# Negative Energy in String Theory and Cosmic Censorship Violation

Thomas Hertog[1], Gary T. Horowitz[1], and Kengo Maeda[2]

[1] *Department of Physics, UCSB, Santa Barbara, CA 93106*
hertog@vulcan.physics.ucsb.edu, gary@physics.ucsb.edu

[2] *Yukawa Institute for Theoretical Physics, Kyoto University, Kyoto 606-8502, Japan*
kmaeda@yukawa.kyoto-u.ac.jp

### Abstract

We find asymptotically anti de Sitter solutions in $\mathcal{N}=8$ supergravity which have negative total energy. This is possible since the boundary conditions required for the positive energy theorem are stronger than those required for finite mass (and allowed by string theory). But stability of the anti de Sitter vacuum is still ensured by the positivity of a modified energy, which includes an extra surface term. Some of the negative energy solutions describe classical evolution of nonsingular initial data to naked singularities. Since there is an open set of such solutions, cosmic censorship is violated generically in supergravity. Using the dual field theory description, we argue that these naked singularities will be resolved in the full string theory.

# 1 Introduction

Anti de Sitter space (AdS) has the property that a scalar field with negative mass squared does not cause an instability provided that $m^2 \geq m_{BF}^2$ where $m_{BF}^2 < 0$ is a certain lower bound known as the Breitenlohner-Freedman (BF) bound [1, 2]. This is important since many supergravity theories arising in the low energy limit of string theory contain fields with negative $m^2$, but they all satisfy this bound. It is commonly believed that there is a positive energy theorem [3, 4, 5, 6] which ensures that the total energy cannot be negative whenever this condition is satisfied.

We will show that, while this is indeed true for $m^2 > m_{BF}^2$, the positive energy theorem *can* be violated for fields which saturate the BF bound. This will be demonstrated by explicitly constructing nonsingular initial data with negative total energy. In fact, the energy can be arbitrarily negative. This is because the positive energy theorem requires boundary conditions which are too strong for fields which saturate the bound. In particular, it requires stronger boundary conditions than those required for finite mass. With the natural boundary conditions, it is only the sum of the usual AdS energy and another (finite) contribution from the asymptotic scalar field which is required to be positive.

The negative energy solutions we find are quite different from previous examples of negative energy in AdS. The AdS soliton [7] has negative energy, but asymptotically approaches AdS only locally. One must introduce a periodic identification asymptotically to construct it. The solutions we present do not require any identification. The counterterm approach can result in negative energy for AdS itself [8]. We will work with the standard energy relative to AdS, so the energy of AdS is zero by definition.

We also consider the evolution of our negative energy configurations and show that an open subset of them lead to naked singularities. This shows that cosmic censorship [9] is violated generically in a low energy limit of string theory. We had previously shown that cosmic censorship can be violated in spacetimes which asymptotically approach AdS [10]. But in that work, we considered gravity coupled to a scalar field with a general potential (satisfying a positive energy theorem), and found that for some potentials cosmic censorship was violated. We did not require that the potential could be derived from string theory. The present work shows that the same phenomena can happen even in string theory.

$\mathcal{N} = 8$ gauged supergravity theories in both four [11] and five [12, 13] dimensions have scalar fields which saturate the BF bound. This is of particular interest since these theories are believed to arise as the low energy limit of string theory (or M theory) with boundary conditions $AdS_4 \times S^7$ [14] or $AdS_5 \times S^5$. For these boundary conditions, we have the powerful AdS/CFT correspondence which relates string theory to a dual field theory [15]. This raises another puzzle since the field theory Hamiltonian is bounded from below, so how can it describe negative energy states in the bulk? The answer seems to be that the field theory Hamiltonian should not be identified with the usual AdS energy, but rather a modified energy which is indeed always positive.

It is of great interest to ask what happens in a full quantum theory of gravity, when the semiclassical solution develops a naked singularity. One can now begin to address this issue using the dual field theory description. Although we have not yet explored this in detail, there appears to be no reason for the field theory evolution to break down. It is likely that the full quantum theory resolves the naked singularity. This provides a new approach for studying cosmological singularities since the naked singularities we find are spacelike inside some region. They are like Big Crunch singularities embedded in an asymptotically anti de Sitter space.

Since boundary conditions will play an important role in our analysis, recall that if we write $AdS_d$ (with unit radius of curvature) in global coordinates

$$ds^2 = -(1 + r^2)dt^2 + \frac{dr^2}{1 + r^2} + r^2 d\Omega_{d-2} \tag{1.1}$$

then solutions to $\nabla^2 \phi - m^2 \phi = 0$ with harmonic time dependence $e^{-i\omega t}$ all fall off asymptotically like $1/r^{\lambda_\pm}$ where

$$\lambda_\pm = \frac{d - 1 \pm \sqrt{(d-1)^2 + 4m^2}}{2} \tag{1.2}$$

The BF bound is

$$m^2_{BF} = -\frac{(d-1)^2}{4}. \tag{1.3}$$

For fields which saturate this bound, $\lambda_+ = \lambda_- \equiv \lambda$ and the second solution asymptotically behaves like $\ln r / r^\lambda$. For definiteness, we will focus on $d = 5$ since the dual field theory is simply four dimensional $\mathcal{N} = 4$ super Yang-Mills. In this case, the fastest that a mode with $m^2 = m^2_{BF}$ can fall off is $1/r^2$.

In the next section, we explicitly construct the negative energy initial data. In section 3, we discuss the positive energy theorem, and the boundary conditions required for it to hold. The following section shows that the negative energy initial data evolve to naked singularities. We also find the corresponding solution of ten dimensional IIB supergravity. Finally, section 5 contains some further discussion, including the implications for cosmological singularities.

## 2   Negative Energy Solutions in Supergravity

$\mathcal{N} = 8$ gauged supergravity in five dimensions [12, 13] is thought to be a consistent truncation of ten dimensional type IIB supergravity on $S^5$. The spectrum of this compactification involves 42 scalars parameterizing the coset $E_{6(6)}/USp(8)$. The fields which saturate the BF bound correspond to a subset that parametrizes the coset $SL(6,R)/SO(6)$. From the higher dimensional viewpoint, these arise from the $\ell = 2$ modes on $S^5$ [16]. The relevant part of the action for our discussion involves five scalars $\alpha_i$ and takes the form [17]

$$S = \int \sqrt{-g} \left[ \frac{1}{2}R - \sum_{i=1}^{5} \frac{1}{2}(\nabla \alpha_i)^2 - V(\alpha_i) \right] \tag{2.1}$$

where we have set $8\pi G = 1$.[1] The potential for the scalars $\alpha_i$ is given in terms of a superpotential $W(\alpha_i)$ via

$$V = \frac{g^2}{4} \sum_{i=1}^{5} \left( \frac{\partial W}{\partial \alpha_i} \right)^2 - \frac{g^2}{3} W^2 \quad , \tag{2.2}$$

$W$ is most simply expressed as

$$W = -\frac{1}{2\sqrt{2}} \sum_{i=1}^{6} e^{2\beta_i} \tag{2.3}$$

---
[1] Our formula's differ slightly from [17], since they use $4\pi G = 1$.

where the $\beta_i$ sum to zero, and are related to the five $\alpha_i$'s with standard kinetic terms as follows [17],

$$\begin{pmatrix} \beta_1 \\ \beta_2 \\ \beta_3 \\ \beta_4 \\ \beta_5 \\ \beta_6 \end{pmatrix} = \begin{pmatrix} 1/2 & 1/2 & 1/2 & 0 & 1/2\sqrt{3} \\ 1/2 & -1/2 & -1/2 & 0 & 1/2\sqrt{3} \\ -1/2 & -1/2 & 1/2 & 0 & 1/2\sqrt{3} \\ -1/2 & 1/2 & -1/2 & 0 & 1/2\sqrt{3} \\ 0 & 0 & 0 & 1/\sqrt{2} & -1/\sqrt{3} \\ 0 & 0 & 0 & -1/\sqrt{2} & -1/\sqrt{3} \end{pmatrix} \begin{pmatrix} \alpha_1 \\ \alpha_2 \\ \alpha_3 \\ \alpha_4 \\ \alpha_5 \end{pmatrix} \qquad (2.4)$$

The potential reaches a negative local maximum when all the scalar fields $\alpha_i$ vanish. This is the maximally supersymmetric AdS state, corresponding to the unperturbed $S^5$ in the type IIB theory. At linear order around the AdS solution, the five scalars each obey the free wave equation with a mass saturating the BF bound. Nonperturbatively, the fields couple to each other and it is generally not consistent to set only some of them to zero. The exception is $\alpha_5$, which does not act as a source for any of the other fields.

We now find a class of negative energy solutions that only involve $\alpha_5$, so $\alpha_i = 0$, $i = 1,..4$ in our solutions. Writing $\alpha_5 = \phi$ and setting $g^2 = 4$ so that the AdS radius is equal to one, the action (2.1) further reduces to

$$S = \int \sqrt{-g} \left[ \frac{1}{2} R - \frac{1}{2} (\nabla \phi)^2 + \left( 2 e^{2\phi/\sqrt{3}} + 4 e^{-\phi/\sqrt{3}} \right) \right] \qquad (2.5)$$

We construct the solutions by first solving the constraint equations on a spacelike surface $\Sigma$. We consider initial data with all time derivatives set to zero. For time symmetric initial data the constraint equations reduce to

$$^{(4)}\mathcal{R} = g^{ij} \phi_{,i} \phi_{,j} + 2V(\phi) \qquad (2.6)$$

For spherically symmetric configurations the spatial metric can be written as

$$ds^2 = \left( 1 - \frac{m(r)}{3\pi^2 r^2} + r^2 \right)^{-1} dr^2 + r^2 d\Omega_3^2. \qquad (2.7)$$

The normalization is chosen so that the total mass is simply the asymptotic value of $m(r)$

$$M = \lim_{r \to \infty} m(r) \qquad (2.8)$$

The constraint (2.6) yields the following equation for $m(r)$

$$m_{,r} + \frac{1}{3}mr(\phi_{,r})^2 = 2\pi^2 r^3 \left[(V(\phi) + 6) + \frac{1}{2}(1+r^2)(\phi_{,r})^2\right] \tag{2.9}$$

The general solution for arbitrary $\phi(r)$ is

$$m(r) = 2\pi^2 \int_0^r e^{-\int_{\tilde{r}}^r \hat{r}(\phi_{,r})^2/3\, d\hat{r}} \left[(V(\phi) + 6) + \frac{1}{2}(1+\tilde{r}^2)(\phi_{,r})^2\right] \tilde{r}^3 d\tilde{r}. \tag{2.10}$$

We now specify initial data for $\phi(r)$ on $\Sigma$. We consider a simple class of configurations with a constant density inside a sphere of radius $R_0$:

$$\phi(r) = \frac{A}{R_0^2} \quad (r \leq R_0), \qquad \phi(r) = \frac{A}{r^2} \quad (r > R_0) \tag{2.11}$$

The fall-off of $\phi$ is motivated as follows. If $\phi \to 0$ slowly, we decrease the contribution to the energy from the positive gradient terms and increase the contribution from the negative potential term. Since we want to try to construct a solution with negative energy, we clearly want $\phi$ to vanish as slowly as possible. It is easy to verify that $1/r^2$ is the slowest fall-off that yields finite total energy. In addition, this behavior is the same as the fall-off of the mode solutions (1.2) of the free wave equation which are going as $\sim 1/r^{\lambda_+}$. One can now easily show that for these initial data, the negative contribution to the mass from the potential more than compensates for the positive contribution from the gradient terms. If we take $0 < A \ll R_0^2$ so that the field is everywhere small then (2.10) gives

$$M \approx -\pi^2 A^2 \tag{2.12}$$

Since we can make $R_0$ and therefore $A$ arbitrarily large, it is clear that the total energy can be arbitrarily negative. For $A > R_0^2$, $\phi$ is not small inside the sphere, but by using the fact that $V(\phi) < -6 - 2\phi^2 - \frac{2}{3\sqrt{3}}\phi^3$ for all $\phi > 0$, one can obtain a general upper limit to the total mass, $M < -\pi^2 A^2/\sqrt{3}$.

We have found that there exist non-singular configurations in $\mathcal{N} = 8$ supergravity with negative total mass. In section 4 we study the evolution of our initial data, but first we explain why this result is not in conflict with the positive energy theorem [4, 6].

# 3 Positive energy theorem

How are our negative energy solutions compatible with the fact that there is a positive energy theorem for supergravity? How are they compatible with the AdS/CFT correspondence since the gauge theory Hamiltonian is bounded from below? To answer these questions, we first review the argument for positive energy of test fields originally given in [1, 2], and then discuss the full nonlinear proof of the positive energy theorem.

## 3.1 Positive energy for test fields

Consider a test field of mass $m^2 = -4$ which saturates the BF bound in $AdS_5$. We start with the action

$$\begin{aligned} S &= \frac{1}{2} \int [-(\nabla \phi)^2 + 4\phi^2] r^3 dt dr d\Omega_3 \\ &= \frac{1}{2} \int \left[ \frac{\dot\phi^2}{(1+r^2)} - (D\phi)^2 + 4\phi^2 \right] r^3 dt dr d\Omega_3 \end{aligned} \quad (3.1)$$

where $D$ is the spatial derivative on a constant $t$ surface. Since the background is static, one can compute the Hamiltonian in the usual way and obtain

$$E = \frac{1}{2} \int \left[ \frac{\dot\phi^2}{(1+r^2)} + (D\phi)^2 - 4\phi^2 \right] r^3 dr d\Omega_3 \quad (3.2)$$

This energy density is not positive definite due to the negative $m^2$. However, if we write $\phi = \psi/(1+r^2)$, substitute into (3.2) and integrate by parts we obtain

$$E = \frac{1}{2} \int \left[ (\dot\psi)^2 + (1+r^2)(D\psi)^2 + 4\psi^2 \right] \frac{r^3}{(1+r^2)^3} dr d\Omega_3 - \oint_\infty \psi^2 d\Omega_3 \quad (3.3)$$

The volume term is now manifestly positive. The surface term vanishes provided $\psi$ goes to zero asymptotically, which means that $\phi$ falls off *faster* than $1/r^2$. But we are interested in solutions that fall off precisely as $1/r^2$. In this case, the surface term is nonzero and manifestly negative. So there need not be a positive energy theorem and indeed, as we have seen, negative energy solutions can occur. Notice that this is possible only for fields which saturate the BF bound. If $m^2 > -4$, then the total energy of any configuration that falls-off as $1/r^2$ diverges. Finite energy configurations must fall-off faster, so the surface term vanishes and the energy is always positive.

It is the delicate cancellation between the $m^2\phi^2$ term and the gradient term in (3.2) which allows fields with $m^2 = -4$ to have $1/r^2$ fall-off and finite energy.

One might have thought that the reaction to this would be to claim that one has positive energy only for $m^2 > m_{BF}^2$. Instead Breitenlohner and Freedman [1] proposed to include the limiting case $m^2 = m_{BF}^2$ and modify the definition of the energy[2]. In the original papers from the early 1980's, this was described in terms of an "improved stress tensor" which corresponds to adding a $\beta R\phi^2$ term to the Lagrangian. In AdS, $R$ is a constant, so this indeed looks like a mass term for a test field. But as soon as one goes beyond the linearized approximation, adding a term like this changes the theory. In the context we have been considering, $\mathcal{N} = 8$ gauged supergravity in five dimensions, there is no $\beta R\phi^2$ term in the action, so this is not an option.

However, one still has the possibility of adding a surface term to the action (3.1) to get

$$\begin{aligned}\tilde{S} &= \frac{1}{2}\int [\phi\nabla^2\phi + 4\phi^2]r^3 dtdrd\Omega_3 \\ &= S + \frac{1}{2}\oint \phi\nabla_\mu\phi dS^\mu\end{aligned} \quad (3.4)$$

Now if one derives the Hamiltonian, one finds an extra surface term in the expression for the energy which exactly cancels the surface term in (3.3). This is possible since if $n$ is the unit radial normal to the sphere at infinity, $\phi n \cdot \nabla\phi = -2\phi^2$. So starting with this modified action, the energy is indeed positive.

## 3.2 Nonlinear positive energy theorem

We now turn to the full positive energy theorem for AdS. This is a generalization of the spinorial proof for asymptotically flat spacetimes given by Witten [18]. We will follow the approach in [6]. The boundary conditions needed to apply this proof do not seem to have been clearly spelled out. In AdS, there are no covariantly constant

---

[2]In [2], Mezincescu and Townsend note that a perturbative analysis is not sufficient to prove stability if the bound is saturated. Later, Townsend [6] performed a nonperturbative analysis in spacetimes of arbitrary dimension, following the approach of Boucher [5], in which he claims to establish a positive energy theorem (and stability) even when the bound is saturated. However, as we will discuss in section 3.2, the proof given in [6] does not apply to the usual AdS energy if $m^2 = m_{BF}^2$.

spinors, but there are "supercovariantly" constant spinors $\epsilon_0$ satisfying

$$\nabla_\mu \epsilon_0 + \frac{1}{2}\gamma_\mu \epsilon_0 = 0 \tag{3.5}$$

where $\gamma^\mu$ are the five dimensional gamma matrices. For the theory we are considering (2.5), the scalar potential is derivable from a superpotential $W(\phi)$ via $V = W'^2 - (4/3)W^2$ (2.2). One now defines a modified derivative $\hat{\nabla}_\mu \equiv \nabla_\mu - \frac{1}{3\sqrt{2}}W(\phi)\gamma_\mu$ and the Nester two-form [19]

$$E^{\mu\nu} \equiv \bar{\epsilon}\gamma^{\mu\nu\sigma}\hat{\nabla}_\sigma \epsilon + \text{h.c.} \tag{3.6}$$

where $\gamma^{\mu\nu\sigma} \equiv \gamma^{[\mu}\gamma^\nu\gamma^{\sigma]}$ and $\epsilon$ is an arbitrary spinor that asymptotically approaches $\epsilon_0$.

Let $\Sigma$ be a nonsingular spacelike surface with boundary at infinity, and let $\epsilon$ be a solution to $\gamma^i \hat{\nabla}_i \epsilon = 0$ (where $i$ runs only over directions tangent to $\Sigma$) which asymptotically approaches $\epsilon_0$. Then the integral of $\nabla_\mu E^{\mu\nu}$ over $\Sigma$ is nonnegative, and vanishes if and only if the spacetime is AdS everywhere. (If there is matter in addition to the scalar field, its stress tensor must satisfy the dominant energy condition.) Hence

$$\oint_\infty E_{\mu\nu}dS^{\mu\nu} \geq 0 \tag{3.7}$$

(Note that the volume element picks out the components orthogonal to the three-sphere at infinity.) If $W$ is constant, this reduces to the usual definition of mass in asymptotically AdS spacetimes. However, in our case $W$ is not constant, and for small $\phi$, $\frac{W}{3\sqrt{2}} \approx -1/2 - \phi^2/6$. So there is an additional surface term

$$\oint_\infty \phi^2(\bar{\epsilon}_0 \gamma_{\mu\nu}\epsilon_0)dS^{\mu\nu} \tag{3.8}$$

Since the area of the $S^3$ at infinity grows like $r^3$, $\phi \sim 1/r^2$, and $\epsilon_0$ is supercovariantly constant at infinity, one might have thought that this surface term would always vanish. But it doesn't. Supercovariantly constant spinors grow like $r^{1/2}$ in AdS (see e.g. [20]). In retrospect this is not surprising since the square of a supercovariantly constant spinor is a Killing vector, and a timelike Killing vector in AdS has norm proportional to $r$. So in order for this surface term to vanish and recover the usual positive energy theorem, one needs $\phi$ to vanish faster than $1/r^2$ at infinity. We have seen that this boundary condition is too strong for fields which saturate the BF bound. In general dimension $d$, the required boundary condition on $\phi$ in order to apply the positive energy theorem[3] is that $\phi$ must vanish faster than $r^{-(d-1)/2}$. A natural way out of this conundrum is to modify the definition of energy to include the extra surface term (3.8). We have seen that the combination of this with the usual energy cannot be negative and vanishes only for AdS.

Supersymmetry implies that the square of the supercharge should be positive. Although we have not checked it, we believe that the supercharge also has an extra contribution in this case, so that the positive quantity is indeed the entire surface term (3.7). It would be interesting to verify this by extending the work of [21] to $\mathcal{N} = 8$ supergravity. Since the Hamiltonian of the dual field theory must be positive (or at least bounded from below if one includes Casimir energy) it should be identified with this modified energy.

## 4  Evolution and Naked Singularities

In this section, we consider the evolution of the negative energy initial data constructed in section 2, and show that they evolve to naked singularities. But first, we point out another interesting difference between the usual energy and the modified energy, which arises in evolution.

### 4.1  Is the energy time dependent?

For fields behaving as $\phi = A/r^2 + O(1/r^3)$ at large $r$, the usual energy is time dependent if $A$ is a function of $t$. There is a nonzero flux of energy at infinity. The modified energy, on the other hand, is always time independent. To see this, it suffices to consider the linearized theory, since $\phi$ is very small asymptotically. In terms of the conserved stress tensor

$$T_{\mu\nu} = \nabla_\mu \phi \nabla_\nu \phi - \frac{1}{2} g_{\mu\nu} [(\nabla \phi)^2 + 2V(\phi)] \tag{4.1}$$

the usual energy (3.2) is just the integral of $T_{\mu\nu} \xi^\mu$ over a spacelike surface, where $\xi^\mu$ is the timelike Killing field. The local flux of energy at infinity is thus $T_{\mu\nu} \xi^\mu n^\nu$ where

---

[3]In [4] it is incorrectly stated that in $d = 4$, one only needs $\phi$ to vanish faster than $1/r$. It is easy to see this is incorrect by constructing negative energy solutions analogous to those in the previous section which vanish like $r^{-3/2}$.

$n^\nu$ is an asymptotic unit radial vector. Integrating this flux between $t_1$ and $t_2$ yields

$$E(t_2) - E(t_1) = \lim_{r\to\infty} \int_{t_1}^{t_2} r^4 dt d\Omega_3 \ \dot\phi(r\partial_r)\phi = -\int [A^2(t_2) - A^2(t_1)]d\Omega_3 \quad (4.2)$$

It is now clear that if we add to the definition of the energy a surface term $\int A^2 d\Omega_3$, the modified energy will be time-independent. This is precisely the same surface term which makes the energy positive. If one wants the usual energy to be time independent, one must require that $A$ be independent of time. This can be achieved by imposing boundary conditions at a large but finite $R$ and requiring $\phi = A/R^2$ (with fixed $A$) at this radius. (This is automatically implemented in most numerical evolution schemes.) The radius $R$ is like a cut-off, and in principle should be taken to infinity to obtain the true solution. The fact that the total energy may be time dependent holds only for fields which saturate the BF bound. If $m^2 > m_{BF}^2$, then finite energy requires fields to fall off faster than $1/r^2$ and then the flux always vanishes at infinity.

## 4.2 Cosmic censorship violation

Recall that our initial data consisted of a constant field $\phi = A/R_0^2$ inside a sphere of radius $R_0$. The proper size of this sphere initially is

$$L \approx \int_0^{R_0} \frac{dr}{[1 + (Hr)^2]^{1/2}} \approx H^{-1} \ln R_0 \quad (4.3)$$

where $H^2 = -V(A/R_0^2)/6$. So for large $R_0$ there is a large region $r < R_0$ of constant energy density and we can model the evolution inside its domain of dependence by a $k = -1$ Robertson-Walker universe,

$$ds^2 = -dt^2 + a^2(t)d\sigma^2 \quad (4.4)$$

where $d\sigma^2$ is the metric on the four dimensional unit hyperboloid. The field equations are

$$\frac{\ddot a}{a} = \frac{1}{6}[V(\phi) - \frac{3}{2}\dot\phi^2] \quad (4.5)$$

$$\ddot\phi + \frac{4\dot a}{a}\dot\phi + V_{,\phi} = 0. \quad (4.6)$$

and the constraint equation is

$$\dot a^2 - \frac{a^2}{6}\left[\frac{1}{2}\dot\phi^2 + V(\phi)\right] = 1 \quad (4.7)$$

It is well known that a homogeneous scalar field, rolling down a negative potential, produces a singularity in finite time [22, 23]. The argument is the following. We start with $\phi = A/R_0^2 \ll 1$ and $\dot\phi = 0$, so initially we have

$$\phi(t) = \frac{A}{R_0^2} \cosh 2t , \qquad a(t) = H^{-1} \cos Ht. \tag{4.8}$$

By (4.5), $\ddot a/a$ is always less than its initial value $-H^2$ (which is close to one). So the scale factor must vanish in a time less than $\pi/2H$. Since $\dot\phi \neq 0$, the vanishing of the scale factor causes the energy in the scalar field to diverge, resulting in a curvature singularity. More precisely, after a certain time $T_0$ the potential term in (4.6) is unimportant and the field behaves as $\dot\phi = c/a^4$, where $c$ is a constant. Matching at $T_0$ gives $c \approx A/R_0^2$. From (4.7) it follows that the change in $\phi$ induces a change in the form of the scale factor when $a^2 \dot\phi^2$ is of order one, which occurs when $a^3 \approx c$. Assuming the potential term is negligible compared to the kinetic term (which can be confirmed after the solution is found) (4.7) reduces to $\dot a^2 - c^2/(12a^6) \approx 0$, which implies $a(t) \propto (T_s-t)^{1/4}$ and hence $\phi(t) \propto -\ln(T_s-t)$. Actually, (4.7) also determines the coefficient so that $\phi = -\frac{\sqrt{3}}{2}\ln(T_s - t)$. Since the scalar field diverges, one has a curvature singularity.

Before one can claim that our initial data evolve to a singularity, one must check that the homogeneous approximation is valid all the way to the singularity. This is not completely obvious since the boundary of the domain of dependence is a null surface, and in pure AdS, a radial light ray can travel an infinite distance in finite time. So we need to calculate the proper distance on the initial surface traveled by the inward going radial light ray from the border of the homogeneous region at $r = R_0$ to the singularity. From the Robertson-Walker form of the metric, this is $l = a(0) \int_0^{T_s} dt/a(t)$. In pure AdS the distance $l$ diverges. But, as we have seen, in our case $a(t)$ changes its form near the singularity resulting in finite $l$. If $\phi(R_0) \approx c \ll 1$ then the cutoff on $t$ where $a(t)$ changes its form occurs close to the maximum value $\pi/2H$, yielding $l \propto -\ln c^{1/3} > 1$ (for instance for $c = .01$ one has $l \approx 2.3$). Since the proper distance is proportional to $\ln r$ this implies that the homogeneous approximation is good all the way to the singularity for radii less than $e^{-l}R_0$. This is much smaller than $R_0$ but it can easily be made arbitrarily large by increasing $R_0$ keeping $\phi(R_0)$ fixed. If $\phi(R_0)$ is of order one, then the size of the singular region is $\sim R_0$, for large $R_0$.

If the total mass could not increase, one could easily show that a black hole could not enclose this singularity as follows. If this singularity lies inside a black hole, then we can trace the null geodesic generators of the event horizon back to the initial surface, where they will form a sphere of radius at least $R_s = e^{-2l} R_0$ (one factor of $l$ is for the reduced size of the domain of dependence at the singularity and the second is because the event horizon is an outgoing null surface (see Fig.1)). The area theorem for black holes only requires the null convergence condition and hence still holds even in theories with $V(\phi) < 0$. Since the area of the event horizon cannot decrease during evolution and we are assuming the mass cannot increase, the initial mass $M$ must be large enough to support a static black hole of size $R_s$. Clearly it is impossible to produce a Schwarzschild AdS black hole, since this requires a positive mass $M_{BH} \propto R_s^4$, and our mass is negative. One could imagine the formation of a black hole with scalar hair, with $\phi(r) \sim r^{-2}$ at large $r$ so that the hair renders the total mass finite and negative. However we have numerically verified that with our potential all black hole solutions with scalar hair have $\phi(r) \sim \ln r/r^2$ at large $r$. Thus our finite mass initial data can not evolve to a black hole with scalar hair.

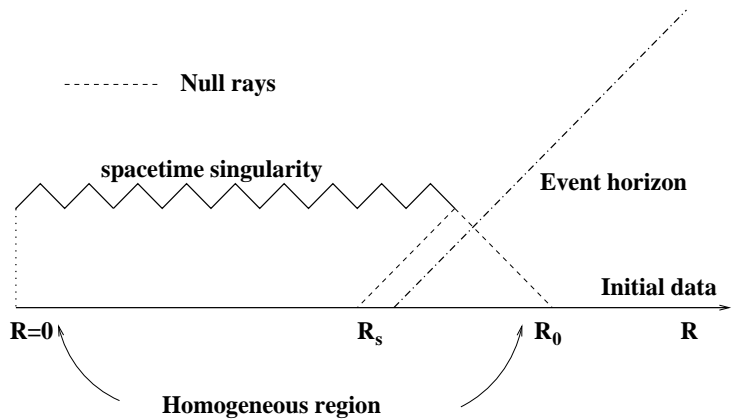

Figure 1: If an event horizon encloses the singularity, it must have an initial size greater than $R_s$.

We have seen that the total mass is not conserved, so it might increase during evolution. If it increases enough, a black hole could form. To ensure that a naked singularity is produced, we can impose a large radius cut-off as mentioned above. This is discussed in more detail in the Appendix. Since the cut-off can be at an arbitrarily large radius, we will continue the following discussion ignoring the cut-off.

Inside the domain of dependence of the homogeneous region the singularity will be spacelike, like a big crunch. The singularity is likely to extend somewhat outside this domain of dependence (so our estimate for $R_s$ is really a lower limit), but not reach infinity. So the singularity will either end or become timelike. In both cases, one has naked singularities. In fact, there is really no way to distinguish these two cases since the evolution ends at the first moment that a naked singularity appears. To see a timelike singularity, one would have to know the appropriate boundary conditions to impose at the singularity, which is not possible classically. If the singularity did reach infinity, it would cut off all space, producing a disaster much worse than naked singularities. But this is unlikely since there would then be a radius $R_c$ on the initial surface such that the outgoing null surfaces for $r > R_c$ expand indefinitely and reach infinity, while those with $r < R_c$ hit the singularity and (probably) contract to a point. This indicates that the surface with $r = R_c$ would reach a finite radius asymptotically, just like the stationary horizons which are ruled out.[4] A similar argument allows us to say something about the geometry near the naked singularity. Consider the area of the spherical cross-sections on an outgoing null surface which hits the naked singularity. If the areas shrink to zero as one reaches the singularity, then a nearby null surface starting at slightly larger radius will have the areas decrease near the naked singularity and then increase as the surface reaches infinity. This contradicts the Raychaudhuri equation and the null convergence condition. We conclude that the area of spheres near the naked singularity remain of nonzero size. The naked singularity is metrically a sphere and not a point. We have seen that inside the domain of dependence of the homogeneous region, the singularity is a strong curvature singularity and all spatial distances shrink to zero. This shows that as the singularity extends outside this region, it becomes weaker, and when it ends, the two sphere remains a finite size. The curvature, however, still diverges.

The above arguments assumed spherical symmetry, but that was not essential. In the central region, the collapsing Robertson-Walker metric develops trapped surfaces. We can clearly perturb our initial data and construct nearby initial data (which need not be time-symmetric) which will still produce trapped surfaces. The singularity theorem guarantees that a singularity must form. On the other hand, the energy will still remain negative, so the singularity cannot be enclosed inside a black hole. Thus cosmic censorship is generically violated in the theory (2.5). In fact, cosmic censorship is generically violated in $D = 5$, $\mathcal{N} = 8$ supergravity, since one can also perturb the other fields in the theory and still produce naked singularities.

---

[4]There is also the possibility that the singularity could become null. If the null singularity reached infinity, it would again cut off all space, and be worse than a naked singularity. If it remained inside a finite region, it would be like a static black hole with singular horizon. Numerical evidence suggests that even these solutions do not exist.

The fact that the naked singularity is not a point holds even for general, nonspherical solutions. To see this, consider the boundary of the past of infinity in the maximal evolution of our initial data. (We are assuming boundary conditions at infinity, so the fact that infinity is timelike is not a problem for evolution.) This is a null surface which ends on the naked singularity. Standard arguments show that this surface is generated by null geodesics which cannot be converging. So the area of any crosssection increases into the future.[5]

## 4.3 Ten dimensional viewpoint

$D = 5$, $\mathcal{N} = 8$ supergravity is believed to be a consistent truncation of ten dimensional IIB supergravity on $S^5$. This means that it should be possible to lift our five dimensional solution to ten dimensions. At the linearized level, the fields which saturate the BF bound correspond to $\ell = 2$ modes on $S^5$. Since the field diverges at the singularity, one expects that the sphere will become highly squashed.

Even though it is not known how to lift a general solution of $D = 5$, $\mathcal{N} = 8$ supergravity to ten dimensions, this is known for solutions that only involve the metric and scalars that saturate the BF bound [24][6]. So we can immediately write down the ten dimensional analog of the solution that evolves to naked singularities. The ten dimensional solution involves only the metric and the self dual five form. To describe them, we first introduce coordinates on $S^5$ so that the metric on the unit sphere takes the form ($0 \leq \xi \leq \pi/2$)

$$d\Omega_5 = d\xi^2 + \sin^2\xi d\varphi^2 + \cos^2\xi d\Omega_3 \qquad (4.9)$$

Letting $f = e^{\phi/2\sqrt{3}}$ and $\Delta^2 = f^{-2}\sin^2\xi + f\cos^2\xi$, the full ten dimensional metric is

$$ds_{10}^2 = \Delta ds_5^2 + f\Delta d\xi^2 + f^2\Delta^{-1}\sin^2\xi d\varphi^2 + (f\Delta)^{-1}\cos^2\xi d\Omega_3 \qquad (4.10)$$

---
[5]One can view this null surface as a type of event horizon since the points inside cannot communicate with infinity. However this event horizon becomes singular and does not correspond to a standard black hole.

[6]We thank Michael Haack for bringing this paper to our attention.

This metric preserves an $SU(2) \times U(1)$ symmetry of the five sphere. The five form is given by
$$G_5 = -U\epsilon_5 - 3\sin\xi \cos\xi f^{-1} * df \wedge d\xi \tag{4.11}$$
where $\epsilon_5$ and $*$ are the volume form and dual in the five dimensional solution and
$$U = -2(f^2 \cos^2\xi + f^{-1}\sin^2\xi + f^{-1}) \tag{4.12}$$

In the homogeneous region of the asymptotic $AdS_5$ space, the metric can be written in Robertson-Walker form (4.4). Near the singularity we have $a(t) \propto (T_s-t)^{1/4}$ and $\phi(t) = -(\sqrt{3}/2)\ln(T_s - t)$. Therefore, over most of the $S^5$ ($\xi \neq \pi/2$) near the singularity, the metric approaches
$$\begin{aligned}ds^2_{10} &= \cos\xi[-(T_s - t)^{-1/8}dt^2 + (T_s - t)^{3/8}d\sigma^2 + (T_s - t)^{-3/8}d\xi^2 \\ &+ (T_s - t)^{-3/8}\tan^2\xi d\varphi^2 + (T_s - t)^{3/8}d\Omega_3]\end{aligned} \tag{4.13}$$

Introducing a new time coordinate $\eta = (T_s - t)^{15/16}$, this takes a simple Kasner like form
$$\begin{aligned}ds^2_{10} &= \cos\xi[-(16/15)^2 d\eta^2 + \eta^{2/5}d\sigma^2 + \eta^{-2/5}d\xi^2 \\ &+ \eta^{-2/5}\tan^2\xi d\varphi^2 + \eta^{2/5}d\Omega_3]\end{aligned} \tag{4.14}$$

Both the five sphere and the asymptotic anti de Sitter space develop a singularity at the same time.

## 5  Discussion

We have shown there are asymptotically anti de Sitter solutions to $\mathcal{N} = 8$ supergravity that have negative total energy. The reason such solutions can exist is that for fields saturating the BF bound the positive energy theorem requires stronger boundary conditions than those required for finite mass. By contrast, in asymptotically flat space (and in anti de Sitter space provided there are no fields saturating the BF bound), the boundary conditions required for finite mass coincide with those required for the positive energy theorem to hold.

Nevertheless supersymmetry guarantees there exists a modified energy that is always positive. This prevents the anti de Sitter vacuum from decaying through quantum tunneling to a different (regular) state with zero energy. So the existence of solutions with negative energy does not imply the anti de Sitter vacuum is unstable. Classical stability at the linearized level comes from the existence of a complete set of mode solutions that are oscillating in time. It is of course much harder to prove that small nonlinear fluctuations do not develop singularities. In light of our results this now becomes a particularly important issue.

The modified energy consists of the usual energy plus an extra surface term depending on the asymptotic value of the scalar field. To better understand this extra term, we briefly review the situation for fields with masses slightly above the BF bound [25]. For scalars in $AdS_d$ with mass $-\frac{(d-1)^2}{4} < m^2 < -\frac{(d-1)^2}{4} + 1$, there are two complete sets of normalizable modes with fall-off $r^{-\lambda_\pm}$ (1.2). Usually, one assumes that the faster fall-off $\lambda_+$ corresponds to nontrivial states in the dual CFT while the slower fall-off corresponds to modifying the CFT by adding sources. This is supported by the fact that solutions that behave as $r^{-\lambda_-}$ near the boundary have divergent total energy. But we know that the dimension of operators in the dual field theory is just $\lambda$, and the field theory contains operators of dimension less than $(d-1)/2$, so if the AdS/CFT correspondence is correct then there must be two ways to quantize this field so that the roles of the modes are interchanged. This is indeed the case. The usual action $S = -\frac{1}{2}\int (\nabla\phi)^2 + m^2\phi^2$ diverges for modes that fall off like $r^{-\lambda_-}$, so it is only appropriate for the faster fall-off. But we can obtain a finite action by adding the surface term $\frac{1}{2}\oint \phi\nabla\phi$. Furthermore, the energy derived from this action is now finite (and positive), even for the modes which fall off like $r^{-\lambda_-}$. For example, in $AdS_5$, by writing $\phi = \psi/(1+r^2)^{\lambda_-/2}$ one can bring the usual energy to the form

$$
\begin{aligned}
E &= \frac{1}{2}\int \left[(\dot\psi)^2 + (1+r^2)(D\psi)^2 + (4\lambda_- + m^2)\psi^2\right] \frac{r^3}{(1+r^2)^{1+\lambda_-}} dr d\Omega_3 \\
&\quad - \lim_{r\to\infty} r^{4-2\lambda_-} \oint_\infty \psi^2 d\Omega_3
\end{aligned}
\quad (5.1)
$$

Adding the surface term to the action cancels the surface term in this expression for the energy, leaving a manifestly positive and finite result.

Even though the surface term one adds to the action here is exactly analogous to the one we added in (3.4), there is an important difference. When the BF bound is saturated, there is only one quantization of the field consistent with the AdS symmetries [25]. Modes which fall off like $1/r^2$ correspond to nontrivial states and modes which fall off like $\ln r/r^2$ correspond to perturbing the theory. The surface term we add is finite and does not change the quantization of the theory, but just the definition of the energy.[7]

Not surprisingly the existence of negative energy solutions has interesting physical consequences. We have shown that a subset of the negative energy configurations evolve to naked singularities. This means that cosmic censorship is violated generically in $D = 5$, $\mathcal{N} = 8$ supergravity, which is the low energy limit of string theory with $AdS_5 \times S^5$ boundary conditions. With $AdS_4 \times S^7$ boundary conditions, one obtains $D = 4$, $\mathcal{N} = 8$ supergravity. Although we have focused on five dimensions, this theory also has fields which saturate the BF bound. So it will also have negative energy solutions and violate cosmic censorship.

It is an open question whether the cosmic censorship hypothesis holds in string theory with asymptotically flat boundary conditions. We have recently shown that a large class of Calabi-Yau compactifications $M_4 \times K$ contain four dimensional potentials which become negative [27]. In this case the positive energy theorem guarantees the positivity of the ADM energy for all solutions that tend asymptotically to the supersymmetric vacuum. However one might imagine there exist certain configurations that have a large central region with negative energy density which develops a singularity, but with a positive total mass that is too small to enclose the singular region by an event horizon. In [10] we have shown this is possible in certain theories with a positive energy theorem in asymptotic AdS space, and we have outlined a (numerical) program to test this possibility in asymptotically flat space.

Our results also provide a new approach for studying cosmological singularities in string theory. The naked singularities we find are spacelike inside some region. They are like Big Crunch singularities embedded in an asymptotically anti de Sitter space. Actually, since our initial data are time symmetric, there are similar singularities in the past. Thus the central regions of our solutions look like (part of) universes beginning in a Big Bang singularity and ending in a Big Crunch (see Fig. 2). It is of great interest to ask what happens to these singularities in a full quantum theory of gravity. One can now begin to address this issue using the dual field theory description. Although we have not yet explored this in detail, we can make the following preliminary comments.

---

[7]If one considers a patch of anti de Sitter space in Poincare coordinates then there is a continuum of oscillating modes with fall-off $1/r^2$ that form a complete set. However, if one considers the modes with fall-off $\ln r/r^2$ the continuum is not complete. There is one extra bound state that grows exponentially in time [26]. Therefore turning on this mode should correspond to an unstable deformation of the boundary theory.

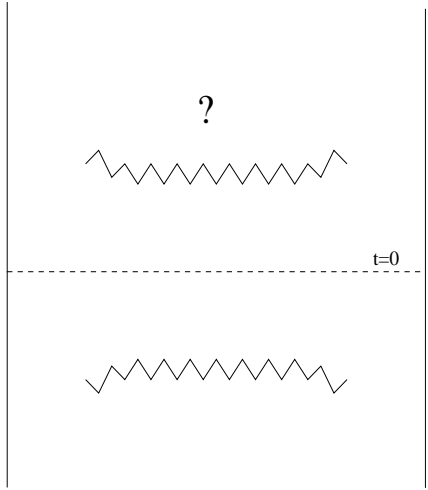

Figure 2: Our solutions are like homogeneous universes beginning in a Big Bang singularity and ending in a Big Crunch, embedded in an asymptotically anti de Sitter space. The dual field description can be used to study how the singularities are resolved.

The scalars which saturate the BF bound in AdS correspond in the gauge theory to the operators $Tr[X^i X^j - (1/6)\delta^{ij} X^2]$ where $X^i$ are the six scalars in $\mathcal{N} = 4$ super Yang-Mills. Each quantum of the field in AdS corresponds to the gauge theory state obtained by acting with one of these operators on the vacuum.[8] To reproduce our classical configuration $\phi(r)$, one could take the corresponding coherent state and map it to the gauge theory. Since this is a finite energy state, there appears to be no reason for the gauge theory evolution to break down. This means that the AdS/CFT correspondence implies that string theory must resolve the naked singularities. In fact one expects even the large N gauge theory to have a well defined evolution for all time, indicating that *classical* string theory should resolve these singularities[9]. This can be viewed as saying that the $\alpha'$ corrections prevent the curvature from diverging, but another interpretation is simply that the spacetime metric is not well defined near

---

[8]For a detailed discussion of the relation between the bulk field and the Yang-Mills operator, see [28].

[9]We thank D. Gross for pointing this out.

the singularity.

Since the singularities are spacelike in the central region, the resolution of the singularity in the gauge theory should determine whether the the universe can "bounce". There are two possibilities. After the formation of the singularity in AdS, the field theory state could correspond to a bulk metric which is semiclassically well defined only outside a finite region, or it could correspond to a metric which is well defined everywhere. In the first case, the classical naked singularity would continue for a while. (It might continue for all time, or eventually the metric could become well defined again and the naked singularity would disappear.) In the second case, the naked singularity would only last an instant. This would be analogous to passing through cosmological singularities in quantum gravity. There has been considerable debate recently about this possibility. We now have a new concrete approach for studying this issue.

### Acknowledgments

It is a pleasure to thank the participants at the KITP program on String Cosmology for discussions, especially T. Banks, D. Gross, R. Kallosh, P. Krauss, D. Marolf, and N. Warner. This work was supported in part by NSF grant PHY-0244764, and a Yukawa fellowship.

# A   Naked Singularities in the Theory with a Cut-off

Our argument for the existence of naked singularities in section 4.2 is incomplete since we had to assume that the total energy did not grow. One can ensure that the usual energy is conserved by picking a large but finite radius $R_1$ and requiring $\phi(R_1)$ to be constant in time. This is automatically implemented in most numerical evolution skemes, and is a standard regulator in discussions of AdS/CFT. However a subtlety now arises since modes that behave like $\ln r/r^2$ near the cut-off cannot be excluded if they have a sufficiently small coefficient. Since these modes fall off more slowly than the usual $1/r^2$ modes we have been considering, their energy is even more negative. The easiest way to show that naked singularities can be produced is to modify our initial data. We consider the following class of configurations,

$$\phi(r) = \phi_0 = \frac{A}{\ln R_1} \frac{\ln R_0}{R_0^2} \qquad (r \leq R_0)$$

$$\phi(r) = \frac{A}{\ln R_1} \frac{\ln r}{r^2} \qquad (R_0 < r < R_1) \tag{A.1}$$

Note that at the cut-off, $\phi(R_1) = A/R_1^2$. For $\phi_0 \ll 1$ and $\frac{\ln R_0}{\ln R_1} \ll 1$ the total mass of the initial data is given by

$$M_i \approx -2\pi^2 A^2 + \frac{\pi^2 A^2}{\ln R_1} + \pi^2 A^2 \left(\frac{\ln R_0}{\ln R_1}\right)^2. \tag{A.2}$$

As we showed earlier, since the density is constant inside the sphere of radius $R_0$, the central region will evolve as a collapsing $k = -1$ homogeneous universe and produce a curvature singularity.

If the singularity lies inside a black hole then the size of its event horizon on the initial surface must be at least $R_s \approx \phi_0^{2/3} R_0$. In addition, the boundary condition on the field at $R_1$ implies that the black hole cannot be Schwarzschild-AdS. It must have some scalar hair. At large radii the hair behaves as

$$\phi(r) = \alpha/r^2 + \beta \ln r/r^2 \tag{A.3}$$

For any value $\phi(R_s) \neq 0$, both coefficients $\alpha$ and $\beta$ are nonzero, but to compute a lower bound on the mass of the black hole solution we consider the case in which all the hair is in the logarithmic mode. The boundary condition then requires $\beta = A/\ln R_1$, and assuming $R_s \gg 1$, we obtain the following estimate for the total mass of the black hole[10],

$$M_{BH} \approx -2\pi^2 A^2 + \frac{\pi^2 A^2}{\ln R_1} + 3\pi^2 \phi_0^{8/3} R_0^4 + 2\pi^2 A^2 \left(\frac{\ln R_0}{\ln R_1}\right)^2 \tag{A.4}$$

Comparison with (A.2) yields, for small $\phi_0$,

$$M_{BH} - M_i \approx \pi^2 \phi_0^2 R_0^4 > 0. \tag{A.5}$$

---

[10]Near the horizon the precise radial profile for $\phi(r)$ deviates from (A.3), but numerical computations show that this estimate is accurate, provided of course that $\frac{\ln R_0}{\ln R_1} \ll 1$.

Since the mass of the black hole is greater than the initial mass (and the mass is now conserved), the singular region can not be enclosed by an event horizon. Cosmic censorship is violated.

This example is not ideal since the limit $R_1 \to \infty$ is not straightforward. Also the $\ln r/r^2$ behavior of the configuration suggests that it should be described by a small perturbation of the gauge theory, and not just a nontrivial state. However the large radius cut-off is believed to be equivalent to a UV cut-off in the gauge theory, and everything that happens in the bulk should have an analog in the dual theory. So one can still hope to use the gauge theory to resolve naked singularities in the bulk.